\documentstyle[12pt]{article}
\def\rf#1{(\ref{eq:#1})}
\def\lab#1{\label{eq:#1}}
\def\nn{\nonumber \\}
\newcommand{\beano}{\begin{eqnarray*}}
\newcommand{\enano}{\end{eqnarray*}}
\def\bea{\begin{eqnarray}}
\def\ena{\end{eqnarray}}
\def\be{\begin{equation}}
\def\ee{\end{equation}}
\def\foot#1{\footnotemark\footnotetext{#1}}

\relax
\def\pa{\partial}
\def\ra{\rightarrow}

\def\PRL#1#2#3{{\sl Phys. Rev. Lett.} {\bf#1} (#2) #3}
\def\NPB#1#2#3{{\sl Nucl. Phys.} {\bf B#1} (#2) #3}

\def\CMP#1#2#3{{\sl Commun. Math. Phys.} {\bf #1} (#2) #3}
\def\PRD#1#2#3{{\sl Phys. Rev.} {\bf D#1} (#2) #3}
\def\PRv#1#2#3{{\sl Phys. Rev.} {\bf #1} (#2) #3}

\def\PLB#1#2#3{{\sl Phys. Lett.} {\bf #1B} (#2) #3}
\def\JMP#1#2#3{{\sl J. Math. Phys.} {\bf #1} (#2) #3}

\def\AoP#1#2#3{{\sl Ann. of Phys.} {\bf #1} (#2) #3}

\def\IJMPA#1#2#3{{\sl Int. J. Mod. Phys.} {\bf A#1} (#2) #3}

\def\PHSA#1#2#3{{\sl Physica} {\bf A#1} (#2) #3}

\def\EPJC#1#2#3{{\sl Eur. Phys. J.} {\bf C#1} (#2) #3}
\def\alv{\vec{\alpha}}
\def\bev{\vec{\beta}}
\def\gav{\vec{\gamma}}
\def\dev{\vec{\delta}}
\def\hv{\vec{h}}
\def\muv{\vec{\mu}}
\def\lav{\vec{\lambda}}
\begin{document}
%%%%%%%%%%%%%%%%%%%%%%%%%%%%%%%%%%%%%%%%%%%%%%%%%%%%%%%%%%%%
\begin{titlepage}
\vspace{-1cm}
\noindent
\hfill{US-FT/20-99}\\
\vspace{0.0cm}
\hfill{hep-th/9910218}\\
\vspace{0.0cm}
\hfill{October 1999}\\
%\vspace{-0.4cm}
%\hfill{(Revised version)}\\
\phantom{bla}
%\vspace{.1in}
\vfill
\begin{center}
{\large\bf  Integrable Quantum Field Theories}\\
\vspace{0.3cm}
{\large\bf  with Unstable Particles}
\end{center}

\vspace{0.3cm}
\begin{center}
J. Luis Miramontes and C.R.~Fern\'andez-Pousa
\par \vskip .1in \noindent
{\em 
Departamento de F\'\i sica de Part\'\i culas,\\
Facultad de F\'\i sica\\
Universidad de Santiago de Compostela\\
E-15706 Santiago de Compostela, Spain}\\
\par \vskip .1in \noindent
e-mail: miramont@fpaxp1.usc.es
\normalsize
\end{center}
\vspace{.2in}
\begin{abstract}
\vspace{.3 cm}
\small
\par \vskip .1in \noindent

\noindent
A new family of $S$-matrix theories with resonance poles is
constructed and conjectured to correspond to the Homogeneous
sine-Gordon theories associated with simply laced compact Lie groups,
where some of the resonance poles can be traced to the presence of
unstable particles in the spectrum. These theories are unitary in the
usual $S S^\dagger =1$ sense, they are not parity invariant, and
they exhibit continuous coupling constants that determine both the mass
spectrum of stable particles and the masses and the position of the
resonance poles. 

\end{abstract}
\vfill
\end{titlepage}

Despite the fact that almost all known particles are
unstable~\cite{Data}, their quantum description is still a
controversial topic, mainly in the relativistic domain~\cite{Gamow}. The
phenomenological relevance of this controversy is exemplified by the
analysis of the LEP and SLC data of electron-positron collisions to study
the (unstable) $Z_0$ boson, which motivated a debate about the proper
definition of the mass of the $Z_0$ leading to theoretical uncertainties
even larger than the experimental error~\cite{ZDebate}. Similarly, the
more convenient phenomenological description of hadron resonances has also
been recently a matter of discussion~\cite{HDebate}. In realistic quantum
field theories, questions like this can generally only be addressed in
perturbation theory or by means of other approximate methods, and so it
would be extremely useful to have simple solvable models whose spectrum
contains unstable particles. 

The construction of solvable models that capture realistic properties of
quantum particles is one of the classical motivations for the study of 
two-dimensional integrable quantum field theories. However, the vast
majority of the integrable models considered so far lack the feature of
including unstable particles. Exceptions are provided by the integrable
perturbation of the
$SO(3)_k$ Wess-Zumino-Witten model discussed by Brazhnikov~\cite{BRA} and,
mainly, by the unitary quantum field theories associated with the
non-abelian affine Toda equations recently constructed by the authors in
collaboration with Gallas and Hollowood in~\cite{MASS}; namely, the
Homogeneous and Symmetric Space sine-Gordon theories~\foot{Using the
construction of ref.~\cite{MASS}, the model of Brazhnikov can be
described as a Symmetric Space sine-Gordon (SSSG) theory associated with
the compact type-I symmetric space $SU(3)/SO(3)$, and it is expected that
many other SSSG theories exhibit similar properties~\cite{Olalla}.}. The
semi-classical spectrum of these models consists of a finite number of
solitonic particles attached to the positive roots of a Lie algebra, but
only those associated with the simple roots turn out to be stable. The
aim of this paper is to propose an exact solution for the Homogeneous
sine-Gordon theories (HSG) corresponding to simply laced Lie groups. 

The exact solution of a two-dimensional relativistic integrable quantum 
field theory is encoded into its two-particle scattering amplitudes.
Indeed, the existence of an infinite number of higher-spin conserved
charges makes the
$S$-matrix factorise into products of two-particle amplitudes. Then, the 
property of factorisation itself and the usual axioms of
$S$-matrix theory constrain the allowed form of the $S$-matrix amplitudes to
such an  extent that it becomes possible to conjecture their
form~\cite{SMAT}. Although it is very  difficult to distinguish physically
between an unstable particle of long lifetime and a stable particle, 
axiomatic $S$-matrix theory makes a big distinction between them. The reason
is that it is based on asymptotic states that exist for arbitrarily long time
and, hence, they can only contain stable particles. In contrast, the basic
property of a long-lived unstable particle, and the one by means of which it
is usually observed, it that it correspond to a `resonance' in interactions
among the stable particles. Therefore, if two stable particles scatter at
centre-of-mass energy $\sqrt{s}$ close to the mass of an unstable state with
appropriate quantum numbers, the corresponding $S$-matrix amplitude is
expected to exhibit a complex pole at $s_R= (M -i\Gamma/2)^2 $ in the
second Riemann sheet, {\it i.e.\/}, off but nearby the real axis in the
complex $s$-plane~\cite{Books}. The position of
the pole is given by the mass, $M$, and decay width, $\Gamma$, of the
resonance, whose inverse is the lifetime of the unstable particle: $\tau=
\hbar/ \Gamma$. If the lifetime is long or, equivalently, if 
$\Gamma \ll M$, the form of the pole is given by the Breit-Wigner 
resonance formula~\cite{Books}:
\be
S \simeq 1- i{2M\Gamma \over s-M^2 +iM\Gamma}
\simeq 1- i{2M\Gamma \over s-(M -i{\Gamma/2})^2} \>.
\lab{BW}
\ee
In a generic theory, unstable particles can also induce other
singularities like complex thresholds and cuts. The simplest would
correspond to the production of the unstable particle together with one
stable particle. However, since the unstable particle decays, this kind
of process would lead to particle production and, hence, it has to be
absent if the theory is integrable. Therefore, we will assume that
resonance poles are the only trace of unstable particles in an
integrable quantum field theory.

So far, only a few examples of factorised $S$-matrix theories with
resonance poles have been considered in the literature. The most
important are the `staircase models', introduced originally by
Al.~Zamolodchikov~\cite{StairZ} and generalized by Martins, Dorey and
Ravanini~\cite{StairD}, where the presence of resonances induces a
remarkable pattern of roaming Renormalization Group trajectories. In
addition, A.~Zamolodchikov and, very recently, Mussardo and Penati have
also proposed interesting factorised
$S$-matrix theories with an infinite number of resonance poles~\cite{MUSS}.
However, although they are completely consistent from the scattering theory
point of view, it is not known whether any of these models allows for a
Lagrangian description, which makes difficult the identification of
resonances with unstable particles.

In contrast, the Homogeneous sine-Gordon theories
always have a unitary Lagrangian description. There is a different HSG
theory for each simple compact Lie group $G$ that corresponds to a unitary
integrable perturbation of the conformal field theory (CFT) associated with
the coset
$G_k/U(1)^{\times r_g}$, where $r_g$ is the rank of $G$, or, equivalently, of
the theory of level-$k$ $G$-parafermions~\cite{PARA}. It is  given by the
action
\be
S_{\rm HSG}[h, A_\pm]\>=\> {1\over \beta^2} \biggl\{ S_{\rm WZW}[h,A_\pm] 
\>-\>\int d^2 x \>V(h) \biggl\}\>,
\lab{Act}
\ee
where $h$ is a bosonic field that takes values in $G$, and $A_\pm$ are
(non-dynamical) abelian gauge connections taking values in a Cartan
subalgebra of $g$, the Lie algebra of $G$. $S_{\rm WZW}[h,A_\pm]$ is the
gauged Wess-Zumino-Witten action corresponding to $G_k/U(1)^{\times
r_g}$, and $\beta^2 = 1/k + O(1/k^2)$ is the coupling constant. Actually, if
the Plank constant is explicitly shown, the last identity becomes 
$\hbar\beta^2 = 1/k + O(1/k^2)$, which exhibits that, just as in the
sine-Gordon theory,  the semi-classical limit is the same as the
weak-coupling limit, and that both are recovered when $k\rightarrow
\infty$. In~\rf{Act}, the potential is $V(h)= - m^2  \langle 
\Lambda_+,  h^{\dagger} \Lambda_- \> h\rangle / \pi$, where $\langle\> ,\>
\rangle$ is the Killing form on $g$, $m^2$ is the only dimensionful constant
of the theory, and $\Lambda_+$ and $\Lambda_-$ are two arbitrary semisimple
elements of the Cartan subalgebra. Assuming that the level $k$ is
larger than the dual  Coxeter number of $G$, it was shown in~\cite{PARA} that
the HSG theories are quantum integrable for any choice of the group $G$ and
for any value of $\Lambda_+$ and $\Lambda_-$ which, therefore, play the role
of coupling constants. $S_{\rm HSG}$ is invariant with respect to the abelian
gauge transformations 
$h(x,t)\mapsto {\rm e\>}^\omega\> h \> {\rm
e\>}^{-\>\tau(\omega)}$ and $A_\pm \mapsto A_\pm \>-
\>\partial_\pm\omega$, 
where $\omega=\omega(x,t)$ takes values in the Cartan subalgebra and
$x_\pm = t\pm x$ are light-cone variables. The precise form of these gauge
transformations is specified by $\tau$, which is an arbitrary element of the
orthogonal group $O(r_g)$ chosen such that $\tau (u) \not= h_{0}^\dagger u
h_0$ for any $u$ in the Cartan subalgebra, where $h_0$ is the 
minimum of the potential. Without loss of generality, we will take
$\Lambda_+$ and $\Lambda_-$ inside  the fundamental Weyl chamber of the 
Cartan subalgebra, such that $h_0=I$ up to gauge transformations.
Moreover, since the spectrum of the HSG theories is independent of $\tau$,
we will only consider the particular choice $\tau=-I$. 

The action $S_{\rm HSG}$ also exhibits an abelian $U(1)^{\times r_g}$ global
symmetry with respect to the group of transformations $ h\mapsto {\rm
e\>}^\alpha\> h\> {\rm e\>}^{-h_{0}^\dagger \alpha h_0}$, where $\alpha$ is
any constant element in the Cartan subalgebra. This implies the existence
of a conserved vector Noether charge. Another important property of the HSG
theories is that they are not parity invariant for generic values of the
coupling constants $\Lambda_+$ and
$\Lambda_-$. Namely, taking into account our choice for $\tau$, they are
not parity invariant unless $\Lambda_+$ is parallel to $\Lambda_-$, a
particular case that will not be considered here. Further
information about these theories can be found in the  original
papers~\cite{MASS,PARA,HSGSOL}. 

The classical integrability of the HSG theories is ensured by the
existence of rank of $G$ conserved quantities for each non-vanishing integer
scale dimension $\pm 1, \pm 2, \ldots$. Their quantum integrability was
established in~\cite{PARA} by explicitly  checking that, after an
appropriate renormalization, all the conserved quantities of spin
$\pm1$ and $\pm2$ remain conserved in the quantum theory. All this
implies that the HSG theories should admit a factorised $S$-matrix
description, and our goal is to construct it starting
from the semi-classical spectrum of the HSG theory associated with a
simply laced  Lie group $G$, which is of solitonic nature. Let us
introduce a  Chevalley basis for the complexification of $g$, the Lie
algebra of $G$, which consists in Cartan generators $\muv\cdot\hv$
together with step operators $E_{\vec
\alpha}$ for each root $\alv$, such that $\muv$ and $\alv$ live in a 
$r_g$-dimensional vector space provided with an inner product normalized 
such that long roots have square length~$2$ (if $g$ is simply laced all
roots  are long). Then, $\Lambda_\pm = \pm i \lav_\pm \cdot \hv$,
and the  condition that $\Lambda_+$ and $\Lambda_-$ are in the
fundamental Weyl chamber translates into $\lav_\pm\cdot \alv >0$ for any
positive root $\alv$ of~$g$. We will also assume that the theory is
not parity invariant, which means that we will exclude the case
when $\lav_+ = \eta \lav_-$ for any real number $\eta$. In~\cite{HSGSOL},
it was shown that the equations-of-motion of the HSG theories admit
time-dependent soliton solutions. In their rest frame, these
solitons provide explicit periodic time-dependent periodic solutions and,
hence, the semi-classical spectrum can be obtained by applying the
Bohr-Sommerfeld quantization rule. For a simply laced group $G$, the
resulting spectrum consists of towers of
$k-1$ soliton particles for each positive root $\alv$ with masses
\be 
M_{\alv} [n] = {k\over \pi}\> m_{\alv}\> \sin\Bigl({\pi n\over k}\Bigr)\>,
\quad n= 1,\ldots, k-1\>,
\lab{MassS}
\ee
where 
\be
m_{\alv} = 2 m\sqrt{(\alv\cdot \lav_+) (\alv\cdot \lav_-)}
\lab{MassF}
\ee
is the mass of the fundamental particle of the theory associated with
$\alv$. As a consequence of the $U(1)^{r_g}$ global symmetry of the theory, these
particles carry a conserved vector Noether charge given by ${\vec Q}_{\alv}
(n) = n \alv$ modulo $k {\Lambda_{R}^\ast}$,
where $\Lambda_{R}^\ast$ is the co-root lattice of $g$, which coincides with
the root lattice if $g$ is simply laced.  Moreover, for a given $\alv$,
the particle labelled by $k-n$ is identified with the anti-particle of the
particle labelled by $n$.

For a fixed root $\alv$, the spectrum is identical to that one of the
minimal $S$-matrix theory associated with the Lie algebra
$a_{k-1}$ or, equivalently, to the
spectrum of the complex sine-Gordon theory~\cite{TN}, which is the HSG
theory associated with $G=SU(2)$. The lightest states associated with a
given root $\alv$ correspond to $n=1$ and $n=k-1$ and have mass $M_{\alv}[1]
= m_{\alv} + O(1/k^2)$ and Noether charge ${\vec Q}_{\alv} [1] = - {\vec
Q}_{\alv} [k-1]=\alv$. These quantum numbers are those of the fundamental
particles of the HSG theory associated with $\alv$ and, hence, these states
are identified with the fundamental particles and anti-particles in the
semi-classical $k\rightarrow \infty$ limit. Moreover, if $n = a+b$ modulo
$k$, one can check that $M_{\alv}[n] < M_{\alv}[a] + M_{\alv}[b]$ and ${\vec
Q}_{\alv} [n]= {\vec Q}_{\alv} [a] + {\vec Q}_{\alv} [b]$,
which leads to the interpretation of the state $(\alv,n)$ as a
bound state of $(\alv,a)$ and $(\alv, b)$, and is the basis of the
bootstrap equations that will be satisfied by our factorised
$S$-matrices. All these identifications can be made because  the solitons
of the HSG theories do not carry topological charges.

Consider now the unique decomposition $\alv = \sum_{ i=1}^{r_g} p_i\>
{\alv}_i$ of a positive root as a linear combination of simple roots
$\alv_1, \ldots, \alv_{r_g}$. It was already pointed out in~\cite{HSGSOL}
that
\be
M_{\alv} [n]\geq \sum_{i=1}^{r_g} M_{\alv_i} (np_i) \quad
{\rm and} \quad
{\vec Q}_{\alv} [n] = \sum_{i=1}^{r_g} {\vec Q}_{\alv_i}[np_i]\>,
\lab{Decay}
\ee
which suggests that the soliton particle $(\alv,n)$ might be unstable and
decay into particles associated with the simple roots. Actually, it can be
easily checked that 
\be
m_{\alv+\bev} \>= \> m_{\alv}\> {\rm e\>}^{\pm \bigl(\sigma_{\alv}\> -\>
\sigma_{\alv+\bev}\bigr)} \>+ \> m_{\bev}\> {\rm e\>}^{\pm
\bigl(\sigma_{\bev}\> -\> \sigma_{\alv+\bev}\bigr)} 
\lab{KinDec}
\ee
where $\alv$, $\bev$, and $\alv+\bev$ are three roots of $g$, and
$ \sigma_{\alv} = {1\over2} \ln {\alv\cdot \lav_+ \over \alv\cdot
\lav_-}$. Eq.~\rf{KinDec} means that the decay of the soliton particle $(\alv
+\bev, n)$ into $(\alv , n)$ and $(\bev, n)$ is kinematically allowed in two
different ways where the outgoing particles would be produced at relative
rapidity $\theta_{\alv} - \theta_{\bev} =  \pm\bigl(\sigma_{\alv} -
\sigma_{\bev}\bigr)$. These two kinematical possibilities for the decay of
the particle $(\alv +\bev, n)$ are related by means of a parity
transformation; however, recall that the theory is not parity invariant
in the generic case we are considering.

To check that this decay is not in contradiction
with classical integrability we have calculated the infinite number of
conserved quantities carried by the soliton particle $(\alv,n)$. This
calculation can be done by using the explicit vertex operator construction
of these soliton solutions presented in~\cite{HSGSOL} together with the
expression for the conserved quantities in terms of tau-functions given
in~\cite{TAU}, and will be presented elsewhere. The result is that the $r_g$
conserved quantities of spin $s$ carried by the soliton $(\alv,n)$
moving with rapidity
$\theta$ are given by the components of
\be
{\vec Q}_{\alv}^{(s)} [n] (\theta) = {\rm e\>}^{s(\theta +
\sigma_{\alv})}\> \sin \Bigl(s\> {\pi n\over k}\Bigr)\> \alv\>.
\ee
They satisfy the following identity
\be
{\vec Q}_{\alv+ \bev}^{(s)} [n] (\theta) \>= \>  {\vec Q}_{\alv}^{(s)}
[n] (\theta- \sigma_{\alv} + \sigma_{\alv+\bev}) \> +\>
{\vec Q}_{\bev}^{(s)}
[n] (\theta- \sigma_{\bev} + \sigma_{\alv+\bev}) \>,
\lab{ClasBoots}
\ee
which, together with~\rf{Decay}, shows that only one of the two
kinematically allowed possibilities for the decay is compatible with
classical integrability; {\it i.e.\/}, when the products move with
relative rapidity
\be
\theta_{\alv} - \theta_{\bev} =  \sigma_{\bev} - \sigma_{\alv} =
\sigma_{\bev, \alv}\>.
\lab{Kin}
\ee

In order to make sure that the decay actually happens in the quantum
theory, we will calculate the decay amplitude and decay
width of the soliton particle $(\alv +\bev, 1)$, which is identified with
a fundamental particle in the semi-classical or weak-coupling
$k\simeq 1/\beta^2\rightarrow \infty$ limit. Then, the decay amplitude can
be calculated in perturbation theory by using the Lagrangian description
of the HSG theories given by the action~\rf{Act}. The tree-level
contribution can be easily obtained by means of the following standard
procedure. First, solve the constraints given by the equations
of motion of the (non-dynamical) abelian gauge connections $A_\pm$.
Second, write the WZW field in terms of complex fields associated
with the fundamental particles,
\be
h(x,t) = \exp \left[ i\sqrt{4\pi\beta^2} \sum_{\alv>0} \Bigl(\phi_{\alv}
E_{\alv} + \phi_{\alv}^\ast E_{-\alv}\Bigr) \right]\>.
\lab{Expand}
\ee
Finally, expand the field $\phi_{\alv}$ in terms of bosonic creation and
annihilation operators for the fundamental particle, $a_{\alv}(k)$, and
anti-particle, $b_{\alv}(k)$,
\be
\phi_{\alv} (x,t) = \int {d k^1 \over 2\pi} {1\over 2k^0} \Bigl( a_{\alv}(k)
{\rm e\>}^{-ik_\mu x^\mu} + b_{\alv}^\dagger(k)
{\rm e\>}^{+ik_\mu x^\mu}\Bigr)\>,
\ee
with the standard relativistic normalization
$[a_{\alv}(k), a_{\bev}^\dagger(q)]\> =\> (2\pi)\> 2k^0\> \delta(k^1 -q^1)
\> \delta_{\alv, \bev}\>$.
Notice that~\rf{Expand} implies a choice for the required gauge fixing
prescription, but the corresponding Fadeev-Popov determinant does not
contribute at tree-level and will not be considered here. The resulting
expansion of the action~\rf{Act} up to order $\beta^2\simeq 1/k$ is
\be
S_{\rm HSG} = \int d^2 x \left[ {\cal L}_2 + {\cal L}_3 + {\cal L}_4 +
\cdots\right]\>,
\lab{SExpand}
\ee
where ${\cal L}_2$ is the kinetic term of the fundamental
particles and anti-particles, 
\be
{\cal L}_2 = \sum_{\alv>0} \left[ \pa_\mu \phi_{\alv} \pa^\mu
\phi_{\alv}^\ast - m_{\alv}^2 \phi_{\alv} \phi_{\alv}^\ast \right]\>,
\ee
while ${\cal L}_3$ and ${\cal L}_4$ provide interaction terms
\bea
{\cal L}_3 = -i{\sqrt{4\pi\beta^2}\over 3!} \sum_{\alv, \bev,\gav}
C_{\alv \bev \gav} \Bigl[ \epsilon^{\mu\nu} \pa_\mu \phi_{\alv} \pa_\nu
\phi_{\bev} \phi_{\gav}  + 4 m^2 (\alv\cdot \lav_+) (\gav\cdot \lav_-)
\phi_{\alv} \phi_{\bev} \phi_{\gav} \Bigr] \>, \nn
\noalign{\vskip 0.2truecm}
{\cal L}_4 = {4\pi\beta^2\over 4!} \sum_{\alv, \bev,\gav, \dev}
\Bigl[ K_{\alv \bev \gav \dev} \pa_\mu \phi_{\alv} \phi_{\bev}
\pa^\mu \phi_{\gav} \phi_{\dev}+ 4 m^2 (\alv\cdot \lav_+) (\gav\cdot \lav_-)
D_{\alv \bev \gav \dev} \phi_{\alv} \phi_{\bev} 
\phi_{\gav} \phi_{\dev} \Bigr] \>, 
\lab{Perturb}
\ena
where the sums extend to all the roots of $g$ (positive and negative), and we
have used the notation $\phi_{-\alv} = \phi_{\alv}^\ast$ and taken
$\epsilon_{01}=1$. Moreover,
\bea
&&C_{\alv \bev \gav} = \langle[E_{\alv}, E_{\bev}], E_{\gav} \rangle\>,
\qquad D_{\alv \bev \gav \dev} = \langle[[E_{\alv}, E_{\bev}], E_{\gav}],
E_{\dev}] \rangle \>,\nn
\noalign{\vskip 0.2truecm}
&&K_{\alv \bev \gav \dev} = D_{\alv \bev \gav \dev} - 3 (\alv\cdot \gav)
\delta_{\alv, -\bev} \delta_{\gav, -\dev}\>. 
\ena
Taking into account
$E_{\alv}^\dagger = E_{-\alv}$, it is straightforward to check that
${\cal L}_3$ and ${\cal L}_4$ are real.

Eq.~\rf{Perturb} shows that the fundamental particle corresponding to the
root $\alv +\bev$ actually decays with an amplitude given by
\bea
&&\langle A_{\alv}(p),A_{\bev}(q) \mid S -1 \mid A_{\alv +\bev}(k)\rangle =
(2\pi)^2\> \delta^{(2)} (p+q-k)\> i {\cal M} \nn
\noalign{\vskip 0.2truecm}
&&\qquad \qquad \qquad = \langle 0| a_{\alv}(p) a_{\bev}(q)\> {\rm
e\>}^{iS_{int}}\>  a_{\alv+\bev}^\dagger(k)
| 0\rangle \nn 
\noalign{\vskip 0.2truecm}
&&\qquad \qquad = (2\pi)^2\> \sqrt{4\pi\beta^2}\> \left(
\epsilon_{\mu\nu} p^\mu q^\nu + m_{\alv} m_{\bev} \sinh\sigma_{\alv, \bev}
\right)\> N_{\alv,\bev}\> \delta^{(2)} (p+q-k)\>,
\ena
where $S_{int} = \int d^2x ({\cal L}_3 + \cdots) $, $[E_{\alv}, E_{\bev}
]= N_{\alv,\bev}\> E_{\alv +\bev}$, and $| A_{\alv}(k)\rangle =
a_{\alv}^\dagger(k) |0\rangle$. In two dimensions it is conventional to
use rapidities instead of momenta: $(p^0, p^1) = (m \cosh \theta, m\sinh
\theta)$. Then, the momentum conservation delta function becomes
\bea
&&\delta^{(2)} (p+q-k) = {1\over m_{\alv} m_{\bev} \sinh
|\theta_{\alv}-
\theta_{\bev}|} 
\> \Bigl[ \delta(\theta_{\alv}
-\theta_{\alv+\bev} +\sigma_{\alv, \alv+\bev})\> \delta(\theta_{\bev}
-\theta_{\alv+\bev} +\sigma_{\bev,\alv+\bev}) \nn
\noalign{\vskip 0.2truecm}
&& \qquad\qquad \> +\> \delta(\theta_{\alv} -\theta_{\alv+\bev}
-\sigma_{\alv, \alv+\bev})\> \delta(\theta_{\bev} -\theta_{\alv+\bev}
-\sigma_{\bev, \alv+\bev})\Bigr]\>,
\ena
which exhibits the two kinematical possibilities for the decay indicated by
eq.~\rf{KinDec} with $\theta_{\alv} - \theta_{\bev} = \pm \sigma_{\alv,
\bev}$. However, since $\epsilon_{\mu\nu} p^\mu q^\nu = m_{\alv} m_{\bev}
\> \sinh(\theta_{\bev} - \theta_{\alv})$, the final expression for the decay
amplitude simplifies to
\bea
&&\langle A_{\alv}(p), A_{\bev}(q) \mid S -1 \mid A_{\alv +\bev}(k)\rangle
= 2 (2\pi)^2\>
\sqrt{4\pi \beta^2}\> {\rm sign\/}(\sigma_{\alv, \bev}) \>
N_{\alv,\bev} \nn
\noalign{\vskip 0.2truecm}
&&{\hskip 4.75truecm}
\delta(\theta_{\alv}
-\theta_{\alv+\bev} +\sigma_{\alv, \alv+\bev})\>
\delta(\theta_{\bev} -\theta_{\alv+\bev} +\sigma_{\bev, \alv+\bev})\>,
\lab{DAmplitude}
\ena
which is consistent with eq.~\rf{Kin} and shows than only the decay that is
compatible with classical integrability actually happens in the quantum
theory, {\it i.e.\/}, $\theta_{\alv} -\theta_{\bev} = \sigma_{\bev,\alv}$.
This is also an indication that the decay is not parity symmetric. For
instance, if $\sigma_{\bev}> \sigma_{\alv}$ particle $\alv$ is produced
precisely on the right-hand-side of particle $\bev$, and the other way round
if $\sigma_{\bev}< \sigma_{\alv}$. From~\rf{DAmplitude}, it is
straightforward to calculate the decay width of the fundamental particle
corresponding to $\alv+\bev$ that will be needed in the following. The
result is
\be
\Gamma_{\alv,\bev} = 2\pi\beta^2 \> {m_{\alv} m_{\bev}\over m_{\alv+\bev}}
\> \sinh |\sigma_{\alv, \bev}| + O(\beta^4) \>,
\lab{DecayW}
\ee
where we have used that, in the Chevalley basis, the structure constants
$N_{\alv,\bev}$ of a simply laced Lie algebra equal $\pm1$. 

Eqs.~\rf{SExpand}--\rf{Perturb} also provide the two-particle
scattering amplitudes in the tree-level approximation. Since the HSG theories
are not parity invariant, these amplitudes will be of the form~\cite{HA}
\be
\langle A_{\alv}(p'),A_{\bev}(q') \mid S \mid
A_{\alv}(p),A_{\bev}(q)\rangle = 4(2\pi)^2 \>
\delta(\theta_{\alv}' -\theta_{\alv}) \delta(\theta_{\bev}' -\theta_{\bev}) 
\> {\cal S\/}^{\alv\> \bev} (\theta_{\alv} -\theta_{\bev})\>, 
\ee
with
\be
{\cal S\/}^{\alv\> \bev}(\theta_{\alv} -\theta_{\bev}) =
\Theta(\theta_{\alv} -\theta_{\bev})\> S_{\alv,\bev}(\theta) \> +\>
\Theta(\theta_{\bev} -\theta_{\alv})\> S_{\bev,\alv}(\theta)\>,
\ee
where $\theta= |\theta_{\alv} -\theta_{\bev}|$ and $\Theta(x)$ is the
usual Heaviside function: $\Theta(x)=0$ if $x<0$ and $=1$ if $x>0$.
$S_{\alv,\bev}(\theta)$ and $S_{\bev,\alv}(\theta)$ are the scattering
amplitudes corresponding to the processes where particle $\alv$ initially is
on the left-hand-side and on the right-hand-side of particle
$\bev$, respectively, which are related by means of a parity 
transformation. The analyticity axiom of $S$-matrix theory postulates
that both functions can be continued to complex values of $\theta$ and,
since the theory is integrable, that the resulting functions will be
meromorphic. Moreover, these meromorphic functions have to be related by
the Hermitian analyticity condition~\cite{HA}
\be
S_{\alv,\bev}(\theta) = \Bigl[S_{\bev,\alv}(-\theta^\ast)\Bigr]^\ast\>.
\lab{HermAn}
\ee
For our purposes, we will only need the scattering 
amplitudes among the fundamental particles associated with the simple roots:
\be
S_{\alv_i, \alv_j}(\theta) = 
\cases{1- i\> \pi\beta^2 \coth {1\over2}(\theta
-\sigma_{ji}) + O(\beta^4)  & if $\alv_i +\alv_j$ is a root, \cr
\noalign{\vskip 0.2truecm}
1+ i\>2\pi\beta^2 \coth {1\over2}\theta + O(\beta^4) & if $\alv_i
=\alv_j$, \cr 
\noalign{\vskip 0.2truecm}
1 & otherwise, \cr}
\lab{PertS} 
\ee
where $\sigma_{ij} = \sigma_{\alv_i} - \sigma_{\alv_i} = - \sigma_{ji}$.
Notice that $S_{\alv_i, \alv_j}(\theta) $ satisfies eq.~\rf{HermAn}. 
Moreover, if $\alv_i +\alv_j$ is a root, it exhibits a pole at $\theta =
\sigma_{ji}$. It is just the resonance pole corresponding to the
unstable fundamental particle associated with the root $\alv_i +\alv_j$, and
its position is in precise agreement with eq.~\rf{Kin}.

Our purpose is to construct factorised $S$-matrix theories compatible with
the properties of the HSG theories. Our main assumption will be that the 
exact spectrum of stable particles coincides with the semi-classical
spectrum of  soliton particles associated with the simple roots of the
algebra. Therefore, according to~\rf{MassS}, there will be $k-1$ particles
for each simple root $\alv_i$ with masses
$ M_i [n] = M_{\alv_i}[n]$, where $n= 1,\ldots, k-1$ and $i=1,\ldots, r_g$.
Recall that particle $(\alv_i, n)$ carries a conserved vector Noether
charge together with an infinite number of higher spin conserved charges
which are proportional to $\alv_i$. This implies that particles associated
with different simple roots belong to different multiplets for any value of
the coupling constants $\lav_\pm$. Moreover, the existence of conserved
charges with even scale dimension allows one to distinguish between
particles and anti-particles. Then, even though
$M_i [n]= M_i [k- n]$, particle $(\alv_i, n)$ and its anti-particle
$(\alv_i, k-n)$ have to be also in different multiplets. Consequently,
scattering has to be purely elastic and our factorised $S$-matrix theory
will be diagonal.

Since the spectrum for a fixed root $\alv_i$ is identical to the
spectrum of the minimal $S$-matrix theory associated with the Lie algebra
$a_{k-1}$, and taking into account~\rf{PertS}, we will assume that the
scattering among particles corresponding
to the same root is described by the minimal $a_{k-1}$ $S$-matrix
amplitudes. Namely, the scattering amplitude between particles $(\alv_i,
m)$ and $(\alv_i, n)$ will be given by~\cite{MinS,Corr}
\be
S_{m\alv_i, n\alv_i}(\theta) = S_{n\alv_i, m\alv_i}(\theta)  = 
S_{m,n}^{a_{k-1}} =
\bigl(m+n\bigr)\>
\left[\prod_{j=1}^{{\rm min}\bigl(m,n\bigr) -1}  \bigl(m+n-2j\bigr)
\right]^2\> 
\bigl(\>|m-n|\> \bigr) \>, 
\lab{MinAk}
\ee
where the standard block notation
\be
(x) = {\sinh {1\over2}( \theta +i{\pi x\over
k}) \over \sinh {1\over2}( \theta -i{\pi x\over
k})}
\lab{Block}
\ee
has been borrowed from~\cite{Corr}. When $m=n=1$, it is straightforward to
check that~\rf{MinAk} agrees with~\rf{PertS} in the $k \simeq 1/\beta^2
\rightarrow \infty$ limit.

The other particles in the semi-classical spectrum are assumed to
be unstable. Therefore, in the limit when their lifetime is long, they
should correspond to resonances poles. In the $k \rightarrow
\infty$ limit, the position of the pole is given by eq.~\rf{Kin}.
Consequently, if $\alv_i +\alv_j $ is a root of $g$ we will assume that
the corresponding unstable particle 
$(\alv_i +\alv_j, 1)$ shows up as a complex simple (resonance) pole in
$S_{\alv_i,\alv_j}(\theta)$ or $S_{\alv_j,\alv_i}(\theta)$ depending on
whether $\sigma_{ji}= \sigma_{\alv_j, \alv_i}$ is positive or 
negative, respectively. Let $\theta_R = \mu -i \varphi$ be the position of
the pole. In the complex $s$-plane, it corresponds to 
\be
s_R =(M- i\Gamma/2)^2 \> =\> m_{\alv_i}^2 + m_{\alv_j}^2 + 2\> m_{\alv_i}\>
m_{\alv_j} \cosh  (\mu  -i\varphi)\>,
\ee
which leads to 
\bea
&& M^2 -{\Gamma^2 \over 4} \> =\> m_{\alv_i}^2 + m_{\alv_j}^2 + 2\> 
m_{\alv_i}\> m_{\alv_j}\> 
\cosh \mu \> \cos \varphi \\
\noalign{\vskip 0.2truecm}
&& M\Gamma \> =\> 2\>  m_{\alv_i}\> m_{\alv_j}\> \sinh \mu \>
\sin\varphi\>. 
\ena
In the $k \simeq 1/\beta^2 \rightarrow \infty$ limit, this pole can be
safely associated with a long lived unstable particle of mass $m_{\alv_i
+\alv_j}$ if the following conditions are satisfied:
\bea
&& M^2 \> \sim \> m_{\alv_i +\alv_j}^2 \> =\> m_{\alv_i}^2 + m_{\alv_j}^2 + 
2\> m_{\alv_i}\> m_{\alv_j}\> \cosh \sigma_{ij}\>, \\
\noalign{\vskip 0.2truecm}
&& M\>\Gamma \> \sim \> m_{\alv_i +\alv_j}\> \Gamma_{\alv_i,\alv_j} \> 
=\> 2\> m_{\alv_i}\> m_{\alv_j}\> \sinh |\sigma_{ij}|\> \pi\beta^2 \>,
\ena
together with $\Gamma\ll M$, where we have used eqs.~\rf{KinDec}
and~\rf{DecayW}. The simplest solution is to take $\mu =
|\sigma_{ij}|$ and $\varphi = \pi/k \simeq \pi\beta^2 \rightarrow 0$, 
{\it i.e.\/}, to fix the position of the resonance pole at $\theta_R =
|\sigma_{ij}| - i\pi/k$. 

Taking into account this together with
eqs.~\rf{HermAn} and~\rf{PertS}, our proposal for the scattering
amplitude between the particles $(\alv_i, 1)$ and $(\alv_j, 1)$ is
\be
S_{\alv_i, \alv_j}(\theta) \> =\> \eta_{i,j}\> (-1)_{\sigma_{ji}} \> \not
=\> S_{\alv_j, \alv_i}(\theta) \>,
\lab{New}
\ee
if $\alv_i + \alv_j$ is a root of $g$, and $S_{\alv_i,\alv_j}=1$
otherwise. In~\rf{New}, $\eta_{i,j} = \eta_{j,i}^{-1}$ is an arbitrary
$k$-th root of $-1$, and we have used the block notation
\be
(x)_\sigma = {\sinh {1\over2}(
\theta - \sigma +i {\pi x\over k}) \over \sinh {1\over2}( \theta -
\sigma -i {\pi x\over k})}\>.
\ee
Notice that $S_{\alv_i, \alv_j}(\theta)$ has a simple pole at $\theta =
\sigma_{ji} -i\pi/k$; however, if it is a resonance pole, it has to be
nearby the physical real positive $\theta$-axis, which requires that
$\sigma_{ji}>0$. Only in this case the pole can be understood as the
trace of an unstable particle. In contrast, also if $ \sigma_{ji} =
-\sigma_{ij} >0$, the pole of $S_{\alv_j, \alv_i}(\theta)$ at $\theta =
\sigma_{ij} -i\pi/k$ is just a shadow pole whose existence is required by
the Hermitian analyticity condition. The $S$-matrix amplitudes among
non-fundamental particles are obtained by means of the bootstrap
principle. In this case, the only possible fusions are $(\alv_i,a)+
(\alv_i,b) \rightarrow (\alv_i,c)$ with
$c= a+b$ modulo~$k$, {\it i.e.\/}, those of the minimal $a_{k-1}$
$S$-matrix theory, which occur at the rapidity values $\theta= iu_{ab}$:
\be
u_{ab}\> =\> \cases{{a+b\over k}\> \pi & if $a+b=c\le k$, \cr
\noalign{\vskip 0.2truecm}
{2k-(a+b)\over k}\> \pi & if $a+b=c+k\ge k$.\cr}
\lab{FAngles}
\ee
The resulting amplitudes are
\be
S_{m\alv_i, n\alv_j}(\theta) = \eta_{i,j}^{m n}\>
\prod_{j=0}^{{\rm min}(m,n) -1} \> (-|m-n| -1-2j)_{\sigma_{ji}} \not= 
S_{n\alv_j , m\alv_i}(\theta)  
\lab{NewS}
\ee
if $\alv_i +\alv_j$ is a root, and $=1$ otherwise. It is easy to check
that all the simple poles of $S_{m\alv_i, n\alv_j}(\theta)$ are located
at $\theta= \sigma_{ji} - i\varphi$ with $0<\varphi <\pi$, as required by
the property of causality.

Since our $S$-matrix theory is diagonal, it trivially satisfies the
Yang-Baxter equation. However, in order to show that eqs.~\rf{MinAk}
and~\rf{NewS} define a consistent theory, it is necessary to
check whether they satisfy the following properties~\cite{SMAT}:
\begin{itemize}
\item[(i)] {\it Analyticity.\/} All the two-particle amplitudes
$S_{A,B}(\theta)$ are meromorphic functions of~$\theta$. Moreover, since
the theory is not parity invariant, the two amplitudes
$S_{A,B}(\theta)$ and $S_{B,A}(\theta)$ are related by means of the
Hermitian analyticity condition
\be
S_{A,B}(\theta) = \Bigl[S_{B,A}(-\theta^\ast)\Bigr]^\ast\>,
\lab{HAPlus}\ee
which is the proper formulation of analyticity in the framework of
standard $S$-matrix theory (see~\cite{HA} and the references therein).

\item[(ii)] {\it Unitarity.\/} The amplitudes given by eqs.~\rf{MinAk}
and~\rf{NewS} satisfy 
\be
S_{A,B}(\theta)\> S_{B,A}(- \theta) \> =\> 1\>.
\lab{QGUnit}
\ee
This is the condition that usually plays the role of unitarity in the
context of the quantum group approach to two-dimensional factorised
$S$-matrix theories, but it is well known that it is not
equivalent to physical unitarity ($S S^\dagger = 1$) unless the $S$-matrix
amplitudes exhibit additional symmetries, like parity or time-reversal
invariance~\cite{TW}. Nevertheless, it was shown in~\cite{HA}
that~\rf{QGUnit} is equivalent to physical unitarity if the
$S$-matrix amplitudes exhibit Hermitian analyticity, which means that
our amplitudes also satisfy
\be
S_{A,B}(\theta)\> [S_{A,B}(\theta)]^\ast \> =\> 1\>,
\lab{QGPhys}
\ee
for real values of $\theta$.  This supports the claim that the HSG
theories are unitary integrable quantum field theories.

\item[(iii)] {\it Crossing symmetry.\/} 
Our amplitudes satisfy
\be
S_{A,B}(i\pi\>- \>\theta)\> =\> S_{B,\overline{A}}(\theta)\>,
\lab{Crossing}
\ee
where the charge conjugation operator maps $A=(\alv_i, n)$ into
$\overline{A}=(\alv_i, k-n)$. Eqs.~\rf{QGUnit} and~\rf{Crossing} imply
that our two-particle $S$-matrix amplitudes are $2\pi i$-periodic
functions, despite not being Real analytic.

\item[(iv)] {\it Bootstrap equations.\/} Since the only possible fusions
are $(\alv_j,a)+ (\alv_j,b) \rightarrow (\alv_j,c)$ with
$c= a+b$ modulo~$k$, which occur at the rapidity values $\theta= iu_{ab}$
given by eq.~\rf{FAngles}, the bootstrap equations read
\be
S_{d\alv_i, c \alv_j} (\theta) \>= \> S_{d\alv_i, a\alv_j}
\bigl(\theta - i\> \overline{u}_{a\> \overline{c}} \bigr)\>
S_{d\alv_i, b\alv_j} \bigl(\theta +  i\> 
\overline{u}_{b\> \overline{c}} \bigr)\>,
\lab{Boots}
\ee
where $ \overline{c} = k-c$ and $\overline{u}_{ab} = \pi - u_{ab}$.
Eq.~\rf{Boots} can be easily checked by showing that both sides have 
simple zeroes and poles located at the same values of $\theta$ and the
same asymptotic $\theta\ra \pm \infty$ behaviour.
\end{itemize}

Moreover, in the $k \simeq 1/\beta^2 \rightarrow \infty$ limit,
eq.~\rf{New} or, equivalently,~\rf{NewS} with $m=n=1$ becomes
\be
S_{\alv_i, \alv_j}(\theta) \> \simeq \> \eta_{i,j}\left(
1- i{\pi\over k} \coth {1\over2}(\theta
-\sigma_{ji}) + O({1\over k^2})\right).
\ee
This agrees with the perturbative result given by eq.~\rf{PertS} up to
the presence of the $\theta$-independent phase factor $\eta_{i,j}$ that is
essential to satisfy both the crossing symmetry relation given
by~\rf{Crossing} and the bootstrap equations~\rf{Boots}. To spell this
out, it is enough to consider $\widehat{S}_{m\alv_i, n\alv_j} =
\eta_{i,j}^{-mn}S_{m\alv_i, n\alv_j}$ and check that
$\widehat{S}_{m\alv_i, n\alv_j} (i\pi -\theta) = (-1)^n
\widehat{S}_{n\alv_j, (k-m)\alv_i} (\theta)$, which shows that
$\eta_{i,j} \eta_{j,i} =1$ and $(\eta_{i,j})^k= -1$ are needed in order to
satisfy the crossing symmetry relation. The bootstrap equations lead to the
same requirements.

It is worth pointing out that the phase factors $\eta_{i,j}$ can
alternatively be taken to be $\theta$-dependent if we are ready to relax
the requirement of analyticity. For instance, if
$\eta_{i,j} \ra {\eta_{i,j}}^{\Theta({\rm Re}(\theta) + a_{i,j})}\;
{\varphi_{i,j}}^{\Theta(-{\rm Re}(\theta) - a_{i,j})}$
in~\rf{NewS}, our $S$-matrix amplitudes still satisfy unitarity,
crossing symmetry and the bootstrap equations provided that $(\eta_{i,j})^k =
(\varphi_{i,j})^k =-1$, $\eta_{i,j}\varphi_{j,i} =1$, and
$a_{i,j}=- a_{j,i}$. Moreover, they also satisfy the Hermitian
analyticity condition~\rf{HAPlus}, even though the amplitudes are not
meromorphic functions of $\theta$ any more. A physical motivation
for a choice like this can be found in~\cite{Asym} where non-analyticity
is related to generalized statistics and is used to make the physical
scattering amplitudes have a trivial asymptotic behaviour $S_{AB}
(\theta)\ra 1$ when $\theta\ra \pm\infty$. Actually, in our case it is
reasonable to demand that $S_{m{\alv_i}, n{\alv_j}}$ becomes trivial also
when $|\sigma_{i,j}|\ra \infty$, {\it i.e.}, in the limit when the mass
$m_{\alv_i +\alv_j}$ of  the unstable particle becomes infinite and,
hence, it is expected to  decouple. Both asymptotic conditions can be
implemented by taking $\eta_{i,j}= \varphi_{i,j}^{-1} = {\rm
e\/}^{i\pi/k}$ and $a_{i,j} = -\lambda^2 \sigma_{j,i}$ for some real
number $\lambda$. 

Along the paper, we have restricted ourselves to the case when parity is
broken in the classical theory, which means that $\lav_+ $ and $\lav_-$
are not allowed to be parallel. However, eq.~\rf{NewS} makes perfect
sense also in this case. If $\lav_+ = \eta\lav_-$ for some real number
$\eta$ then $\sigma_{i,j}=0$ for any $i,j$ and, for $i\not=j$, all the
poles of $S_{m\alv_i, n\alv_j}(\theta)$ are located along the negative
part of the imaginary $\theta$-axis, which means that they can be
associated with `virtual states' instead of unstable particles.
Nevertheless, even in this case the factorised $S$-matrix theory fails to
be parity invariant because
$\eta_{i,j}\not=\eta_{j,i}$. Actually, these
$\theta$-independent phase factors are reminiscent of the two-particle
scattering amplitudes of the Federbush model, which is the simplest
non-parity invariant integrable theory~\cite{Feder}.

In conclusion, eqs.~\rf{MinAk} and~\rf{NewS} define consistent
$S$-matrix theories compatible with the semi-classical spectrum of the
simply laced HSG theories. This supports the conjecture that they
actually provide the exact solution of the HSG theories corresponding to
simply laced compact simple Lie groups. Since the HSG
theory associated with a Lie group $G$ is an integrable perturbation of
the theory of level-$k$ $G$-parafermions, our next task is to
probe the UV limit of the proposed $S$-matrix theories using the
thermodynamic Bethe ansatz. In particular, this analysis will give the
value of the central charge of the underlying UV conformal field theories
which, in our case, should be $(k-1)\> h_g r_g/(k+h_g) $, where $r_g$ and
$h_g$ are the rank and Coxeter number of $G$,
respectively. An important novel feature of this analysis, which is already
in progress~\cite{NEW}, is the loss of parity invariance. Moreover, the
existence of different mass scales depending on the values of the coupling
constants
$\lav_+$ and $\lav_-$ is expected to induce a `staircase-like' behaviour of
the effective central charge. This should be a consequence of the change in
the number of light degrees of freedom associated with the decoupling of the
heavy particles, both stable and unstable. It is worth noticing that in the
case of the original staircase models~\cite{StairZ,StairD} such a direct
physical interpretation is not possible due to the lack of a Lagrangian
description.

\vspace{0.75 truecm}

\noindent\centerline{\large\bf Acknowledgments} 

\vspace{0.25truecm}
We would like to thank J.~S\'anchez Guill\'en, A.~Fring, H.~de Vega, and
C.~Korff for valuable discussions. This research is supported partially by
CICYT (AEN96-1673), DGICYT (PB96-0960), and the EC Commission via a TMR
Grant (FMRX-CT96-0012).

\end{document}